\documentclass[aps,prl,twocolumn,showpacs,superscriptaddress]{revtex4-1}
\usepackage{amsmath}
\usepackage{epsfig}
\usepackage{amssymb}
\usepackage{dcolumn}
\usepackage{graphicx}
\usepackage{dcolumn}

\usepackage{color}

\newcommand{\beq}{\begin{equation}}
\newcommand{\eeq}{\end{equation}}

\begin{document}
\date{\today}

\title{On-demand dark soliton train manipulation in a spinor polariton condensate}

\author{F. Pinsker}
\affiliation{Department of Applied Mathematics and Theoretical Physics, University of Cambridge, United Kingdom.}
\email{fp278@cam.ac.uk}

\author{H. Flayac}
\affiliation{Institute of Theoretical Physics, \'{E}cole Polytechnique F\'{e}d\'{e}rale de Lausanne EPFL, CH-1015 Lausanne, Switzerland.}
\email{hugo.flayac@epfl.ch}

\begin{abstract}
We theoretically demonstrate the generation of dark soliton trains in a one-dimensional exciton-polariton condensate within an experimentally trivial scheme. In particular we show that the frequency of the train can be finely tuned fully optically or electrically to provide a stable and efficient output signal modulation. Taking the polarization degree of freedom into account we elucidate the possibility to form on-demand half-soliton trains.
\end{abstract}
\pacs{03.65.-w, 05.45.-a, 67.85.Hj, 03.75.Kk}

\maketitle

\emph{Introduction}.--
The first unambiguous observation of Bose-Einstein condensation in dilute Bose gases at low temperature \cite{first} set off an avalanche of research on this new state of matter. The lowest energy fraction of a degenerated Bose gas occupying low energy modes obeys the property of vanishing viscosity and does not take part in the dissipation of momentum, a phenomenon referred to as superfluidity \cite{London}. This holds true as long as the condensate is only slightly disturbed \cite{evid}. As soon as strong dynamical density modulations occur, e.g. when the condensate is abruptly brought out of its equilibrium through an external perturbation, it responds in a unique way by generating robust elementary excitations such as solitons in $1$D and vortices in $2$D settings \cite{dark}.

More recently the concept of macroscopically populated single particle states \cite{E,OP} was transposed to a variety of mesoscopic systems such as cavity photons \cite{photon,Light}, magnons \cite{MagnonsBEC}, indirect excitons \cite{IndexBEC}, exciton-polaritons (polaritons) \cite{exit} and even classical waves \cite{Fleisch}. In the proper regime all those systems can be described by complex-valued order parameters - the condensate wave functions - with dynamics governed by nonlinear Schr\"odinger-type equations (NSE) such as the Gross-Pitaevskii (GP) \cite{PitaGross} and the complex Ginzburg-Landau equation (cGLE) \cite{Berloff}. Here the nonlinearity associated with self-interactions plays an essential role on the possible states with or without excitations, their dynamics and in particular their stability \cite{World}. Similarly in the slowly varying envelope approximation light waves can be approximated by complex-valued wave functions governed by NSE's that are formally comparable to those of BECs and thus show analog dynamical behavior such as stationary and moving optical dark or bright solitons in quasi 1D settings \cite{classic, cla}. For several decades light waves have been utilized in a wide range of applications such as in nonlinear fibre optic communication \cite{classic, classic2, classic3, classic4} while research on new technologies is thriving in particular on elementary circuit components such as optical diodes \cite{diode}, transistors \cite{trans} or realizations of analog devices involving exciton-polariton condensates \cite{trans22,trans33} and conceptually on optical computing schemes \cite{opco}.

Polaritons are half-light half-matter quasi-particles formed in semiconductor microcavities and allow high speed propagation from their photonic part while having strong self-interaction from their excitonic fraction. They
are extremely promising from both the fundamental and technological point of view given the ease it provides to finely control the parameters of their condensate now routinely observed in different geometries (see e.g. \cite{BB}). Indeed, the actual technology allows to etch any sample shape to sculpt the confining potential seen by the condensate at will. It explains the plethora of recent proposals \cite{GaoPRB2013,THzPRL1,LiewPRL2008,PavlovicPRL2009,JohnePRB2010,EOArX2013,FlayacRouter,PolaritonDevices} for polariton devices some of which have been experimentally implemented \cite{Racquet,trans22,trans33}. The main advantage with respect to standard optical systems in nonlinear media is the very large exciton-mediated nonlinear response of the system reducing the required input power by orders of magnitude. Recently there was a growing interest in demonstrating the formation of (spin polarized) topological defects \cite{V,HV,S,HS,HSS} that are now envisaged as stable information carriers \cite{Monopole1,Monopole2,Magnetricity} within a young field of research called spin-optronics \cite{ReviewSpin}.

In this letter we shall present experimentally trivial schemes for the intended generation and manipulation of stable and fully controllable wave patterns within a quasi-1D microcavity. We will demonstrate the on-demand formation of dark soliton trains within a quasi-1D channel and the optical and electrical dynamical control of their frequency. Finally we will demonstrate the possibility to control the polarization of the soliton trains.

\begin{figure}[ht]
\includegraphics[width=0.45\textwidth,clip]{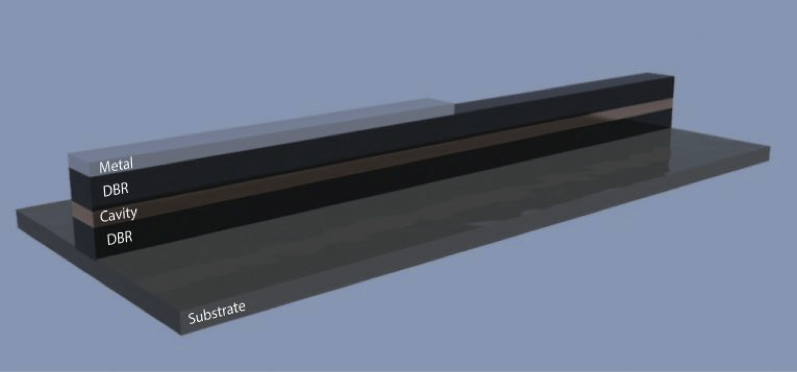}\\
\caption{(Color online) Model of a potential sample consisting in a quasi-1D microcavity (DBR=distributed Bragg reflectors) embedding a metallic deposition over half of its length to form a potential step. A gate voltage can be applied to the metal to tune dynamically the step amplitude.}
\label{Fig1}
\end{figure}

\emph{The model}.--
We shall consider the system modeled in Fig.\ref{Fig1}, namely a wire-shaped microcavity similar to the one implemented in Ref.\cite{Wertz} that bounds the polaritons to a quasi-1D channel. A metallic contact is embed over half of the sample to form a potential step seen by the polaritons and whose amplitude can be tuned on-demand applying an electric field \cite{VortexGun}. The spinor polariton $\boldsymbol{\psi}=(\psi_+,\psi_-)^T$ field evolves along a set of effectively 1D cGLEs coupled to a rate equation for the excitonic reservoir,
\begin{eqnarray}
\nonumber i\hbar \frac{{\partial {\psi _ + }}}{{\partial t}} &=& \left[ { - \frac{{{\hbar ^2}\Delta }}{{2m}} + {\alpha _1}\left( {{{\left| {{\psi _ + }} \right|}^2} + {n_R}} \right) + {\alpha _2}{{\left| {{\psi _ - }}
\label{psi1}
\right|}^2}} \right]{\psi _ + }\\
 &+& \left[ {U - \frac{{i\hbar }}{2}\left( {\Gamma  - \gamma {n_R}} \right)} \right]{\psi _ + } - \frac{{{H_x}}}{2}{\psi _ - }\\
\nonumber i\hbar \frac{{\partial {\psi _ - }}}{{\partial t}} &=& \left[ { - \frac{{{\hbar ^2}\Delta }}{{2m}} + {\alpha _1}\left( {{{\left| {{\psi _ - }} \right|}^2} + {n_R}} \right) + {\alpha _2}{{\left| {{\psi _ + }}
\label{psi2}
\right|}^2}} \right]{\psi _ - }\\
 &+& \left[ {U - \frac{{i\hbar }}{2}\left( {\Gamma  - \gamma {n_R}} \right)} \right]{\psi _ - } - \frac{{{H_x}}}{2}{\psi _ + }\\
\label{nR}
\frac{{\partial {n_R}}}{{\partial t}} &=& P - {\Gamma _R}{n_R} - \gamma \left( {{{\left| {{\psi _ + }} \right|}^2} + {{\left| {{\psi _ - }} \right|}^2}} \right){n_R}.
\end{eqnarray}
This model describes in a simple way the phenomenology of the condensate formation under non-resonant pumping. We assume a parabolic dispersion of polaritons associated with an effective mass $m=5\times10^{-5}m_0$ where $m_0$ is that of the free electron and a decay rate $\Gamma=1/100$ ps$^{-1}$. $U(x,t)=\left[U(t)+U_0\right]{{H}}(x)$, where $H(x)$ is the Heaviside function $U_0=-0.5$ meV is the step height induced by the presence of the metal solely and $U(t)$ is the potential landscape imposed by the external electric field. $\alpha_{1}=1.2\times10^{-3}$ meV$\cdot\mu$m and $\alpha_2=-0.1\alpha_1$ are respectively the parallel and antiparallel spin interaction strength. $H_x=0.01$ meV is the amplitude of the effective magnetic field induced by the energy splitting between TE and TM eigenmodes that couples the spin components \cite{Racquet}. The excitonic reservoir characterized by the decay rate $\Gamma_R=1/400$ ps$^{-1}$ is driven by the pump term $P=A_P \exp(-x^2/\sigma^2)$ where $\sigma=20$ $\mu$m and $A_P$ is taken in the range of hundreds of $\Gamma_R$. It exchanges particles with the polariton condensate at a rate $\gamma=2\times 10^{-2}\Gamma_R$.

We note that while the stimulated scattering is taken into account by Eqs.(\ref{psi1}-\ref{nR}), energy relaxation processes dominant under the pump spot, apart from the lifetime induced decay of the interactions energy, are neglected in this framework and could be treated e.g. within the formalisms of Refs.\cite{LiewTransistor,SavenkoPRL2013}. Energy relaxation would not impact our results qualitatively especially for the finite pump spot size we consider here.

\emph{Soliton train generation}.--
As shown in Ref.\cite{dark} a local abrupt change of self-interaction strength of the condensate leads to the formation of a stable and regular dark soliton train. It happens when the flow in the direction of decreasing interaction due to particle repulsions is locally crossing the speed of sound $c_s(x)=\sqrt{\mu(x)/m}$ where $\mu(x)=\alpha_1 n(x)$ (for a scalar condensate) at the point of abrupt change in self-interactions, solitons are formed from dispersive shock waves \cite{Kamchatnov} that dissipate the local excess of energy. In polariton condensates the interaction strength $\alpha_1$ is varied tuning the exciton/photon detuning and therefore the excitonic fraction, but it can hardly be made inhomogeneous within a given sample nor tuned dynamically. A valuable alternative we follow here is to introduce the tunable potential step $U(x,t)$ in Eqs.(\ref{psi1},\ref{psi2}). The mechanism for soliton generation is the following (see supplemental material for a more complete argument \cite{Supplemental}). Let us suppose we consider the homogeneous density $n_0$ at $t=0$ and neglect the finite lifetime of our quasi particles and the geometry of our pump spot. Then taking the potential $U$ stepwise for all following $t > 0$, we get close to the breaking point at $x=0$ the density $n_1$ for $x=0^-$ and $n_2$ as $x=0^+$ and we say $n_1 = k_1 n_2$ with $1>k_1>0$. Using momentum and mass conservation at $x=0$ we find the simple criterion $0.6404>k_1> 0$ to break the speed of sound \cite{Supplemental}, which is in good agreement with our numerical results. Importantly above this threshold the frequency of the subsequently generated train is increasing with the magnitude of the variation \cite{Supplemental} as the corresponding increase of mass passing the step at $x=0$ allows a more frequent breaking of the local speed of sound. This is analogous to the situation of a superfluid passing an obstacle above criticality for which greater mass transport is equivalent to a higher number of generated vortices in 2D \cite{Berloff2}.

For a given metal type and deposition thickness on top of the microcavity, Tamm plasmon-polariton modes \cite{Tamm} were predicted to form at the interface inducing a local redshift of the polariton resonances of amplitude $U_0$ and the required potential step. The application of a voltage to the metal produces the additional gate redshift $U(t)$ through the excitonic Stark effect up to a few meVs for voltages lying in the range of tens of kV$/$cm \cite{LiewCircuit} and standing for the input modulation of the polariton condensate. The non-resonant excitation of the system is crucial since in this context the condensate phase is free to evolve under the pump spot by contrast to a resonant injection scheme that would imprint the phase preventing the onset of solitons.

\begin{figure*}[ht]
\includegraphics[width=0.9\textwidth,clip]{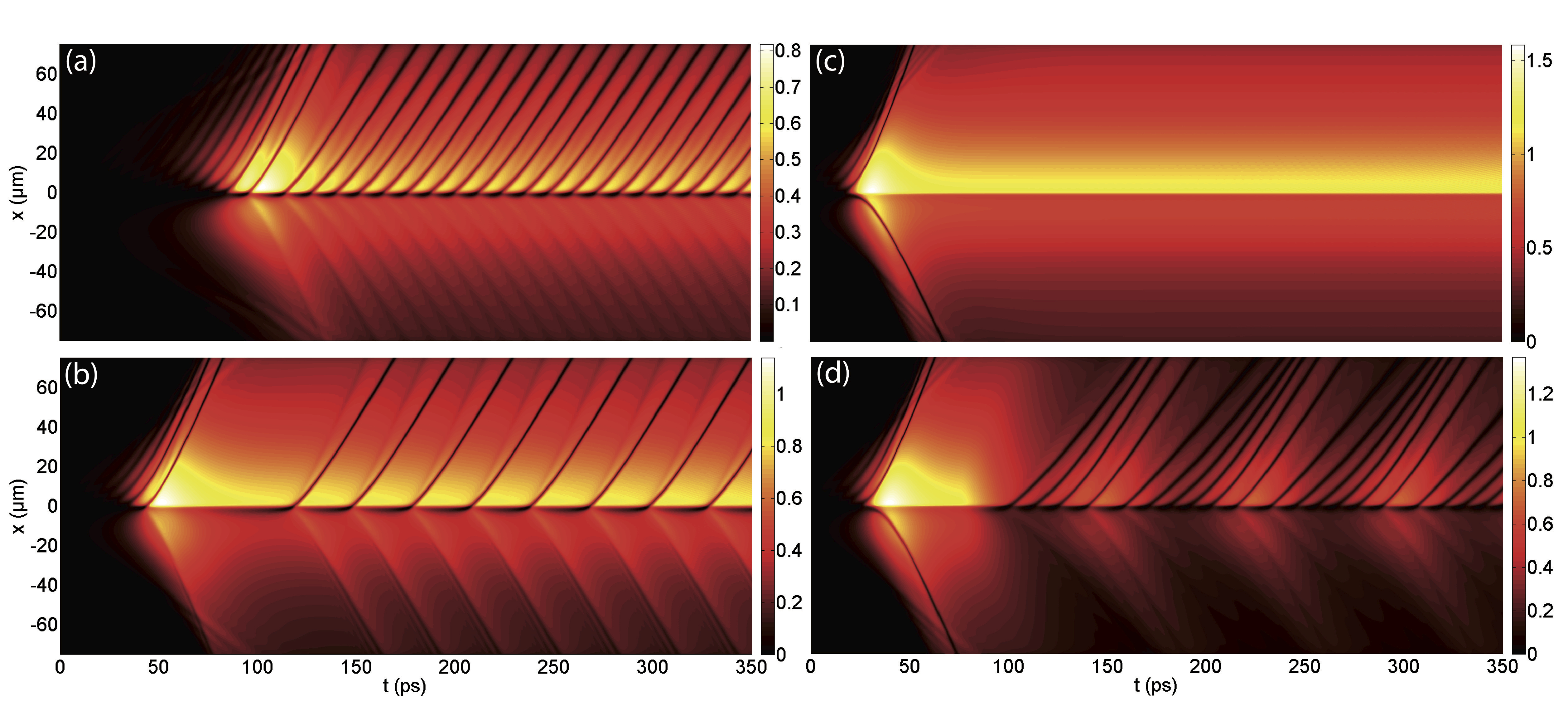}\\
\caption{(Color online) Optical control. (a), (b) (c) Shows results obtained by pumping over the potential step with linearly increasing pump power, we monitor here $\mu(x,t)$ in the colormap. (d) Sinusoidal modulation of the pump power producing a signal frequency modulation.}
\label{Fig2}
\end{figure*}

\emph{Optical control}.--
Let us start with the simplest passive configuration where no voltage is applied and therefore the potential step is \emph{fixed}. We switch on the pump laser focussed on the step at $t=0$ and wait for the steady state to be reached. The reservoir is filled by the incoherent pump and the stimulation towards the lowest polariton energy state occurs forming the condensate with a chemical potential $\mu = (\alpha_1 + \alpha_2)/2 n-H_x/2$ (corresponding to the measurable blueshift of the polariton emission) where $n=|\psi_+|^2+|\psi_-|^2=n_++n_-$ is the total polariton density. Given the interrelation $\alpha_1>\alpha_2$ the condensate interaction energy is minimized for a linear polarization meaning that $n_+=n_-$ and the condensate is said to be antiferromagnetic \cite{FlayacSS}. The linear polarization orientation is homogeneous at zero temperature and fixed by the $H_x$ contribution namely along the axis of the wire. In our model we trigger the condensation on the $x$-polarized ground state with initial populations $n_\pm^0$. Fig.\ref{Fig2} shows numerical solutions to Eqs.(\ref{psi1},\ref{psi2},\ref{nR}). We depict the chemical potential $\mu(x,t)$ for crescent pump amplitudes $A_P$. We clearly see the decrease in the train frequency $\nu$ with increasing pump power [panels (a) and (b)] until the train vanishes [panel(c)].

The condensate heals from the step forming a gray soliton resulting from the local velocity gradient, as it happens e.g. at the boundaries of condensate trapped in a square potential. The depth of the soliton is imposed by the local background density and velocity. For high enough background density, the flow is superfluid ($v<c_s$) both around the step and within the soliton that remains pined to the step preventing the train onset [panel(c)]. For lower densities (pump power), the speed of sound can be surpassed at the soliton core which allows the condensate to dissipate the local excess of energy via a dispersive shock wave \cite{Kamchatnov} that releases the soliton to the side where the background flow is the highest. Then it takes some time for the condensate to form a new soliton. The higher the density the stiffer is the condensate and therefore the more time it takes to form a new density deep which governs the quasi-linear train frequency $\nu$ dependence over the chemical potential shown in the supplemental material \cite{Supplemental}. We note that the additional repulsion brought by the reservoir under the pump spot makes analytical estimation from equilibrium condensate irrelevant.


These results demonstrate the possibility to modulate passively an optical signal via the formation of stable dark solitons varying the pump amplitude. The dark soliton signals shall then be detected experimentally at the output via one of the schemes proposed in the context of nonlinear optics \cite{Detector3} to encode information. Indeed, as proposed in \cite{Pinsker} soliton trains can be used to store numbers determined uniquely by an adjustable $\nu$. So far, most of the device proposals involving microcavity polaritons have focussed on signal transmission but rarely on its modulation. Nonetheless, as one can see, the train frequencies lie in the range of THz allowing to perform very high speed processing due to the polariton photonic part combined with a large exciton-mediated nonlinear response.

In Fig.\ref{Fig2}(d), we show an example of sinusoidal input power modulation that leads to a dynamical variation of $\nu$ or a modulation of the output on-demand to produce useful wavepackets. The main advantage of this all-optical input modulation scheme is that it allows us to reach high speed variation of $\nu$ while the drawback is that the background density of the condensate is obviously affected as well. Finally we note that this setup involving a fixed potential step doesn't require specifically a metallic deposition. A sample split in two parts with slightly different lateral width should be sufficient to reproduce the effects discussed above.

\begin{figure}[ht]
\includegraphics[width=0.45\textwidth,clip]{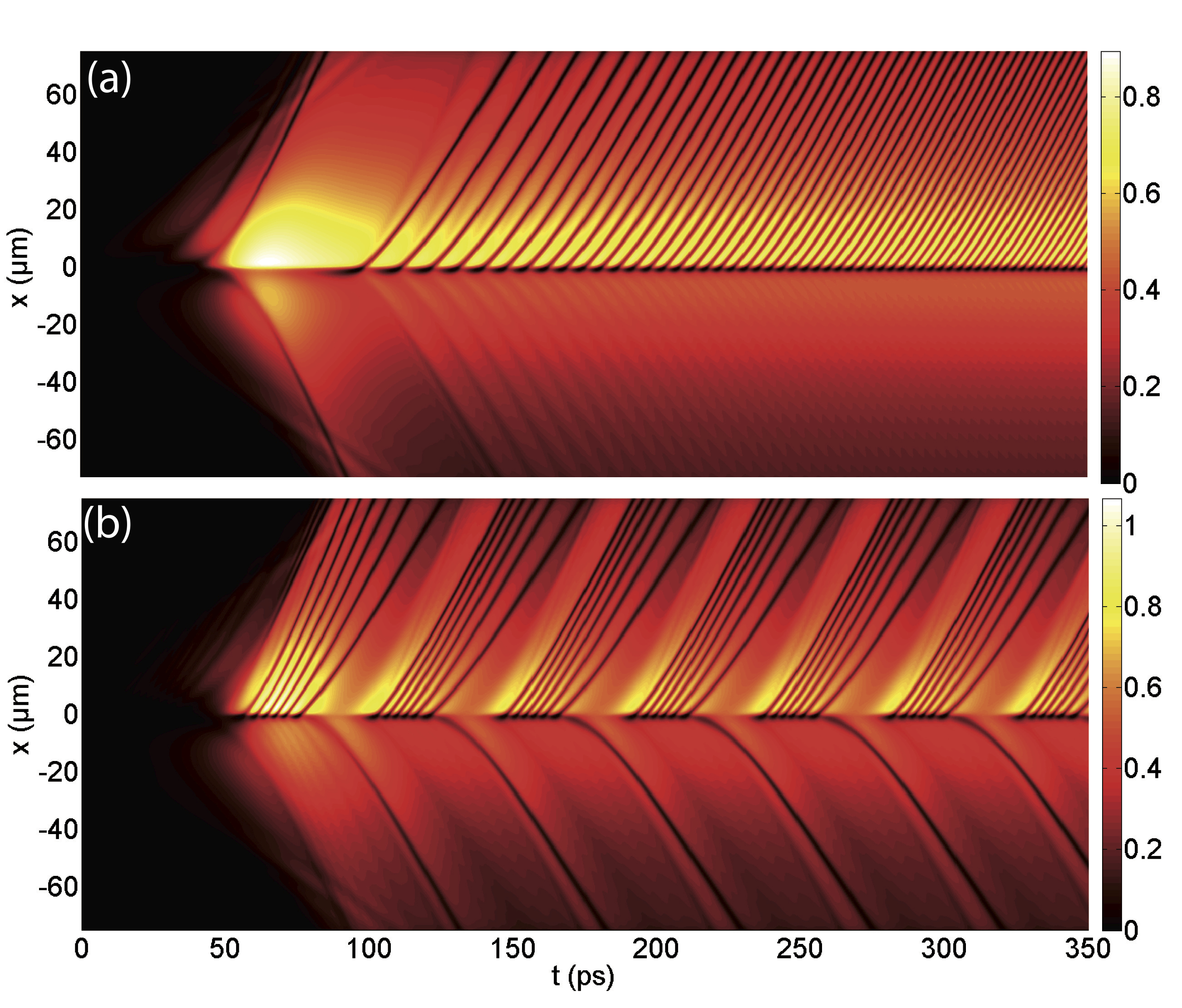}\\
\caption{(Color online) Electric control of the dark soliton trains. (a) The potential step amplitude $U(t)$ is linearly ramped down versus time from 0 meV to -1.5 meV. (b) Sinusoidal modulation of of the step.}
\label{Fig3}
\end{figure}

\emph{Electric control}.--
Now let us consider the case where the pump power is fixed and in addition an electric field is applied to the metallic contact to modulate the potential step height. Under such assumptions, the chemical potential $\mu$ is globally fixed. The higher the step (the electric field), the larger the density gradient and hence one encounters a greater mass transport towards lower energy regions. So similarly to \cite{Berloff2} we obtain an increase in dark solitons train frequency as shown in \cite{Supplemental}. To demonstrate this behavior, we show in the Fig.\ref{Fig3}(a) the results obtained by ramping down linearly the potential step from 0 meV to -1.5 meV which corresponds to an increase in the electric field amplitude. We clearly see the linear increase in $\nu$ versus time.

Similarly to the results of Fig.\ref{Fig2}(d) we show in Fig.\ref{Fig3}(b) results obtained from a sinusoidal modulation of the potential step amplitude producing an efficient dynamical modulation of the output signal in the form of wavepackets. Such an electric control of the polariton flow has the advantage of impacting weakly on the background density but there might be some technological limitation on the switching speed.

\begin{figure}[ht]
\includegraphics[width=0.45\textwidth,clip]{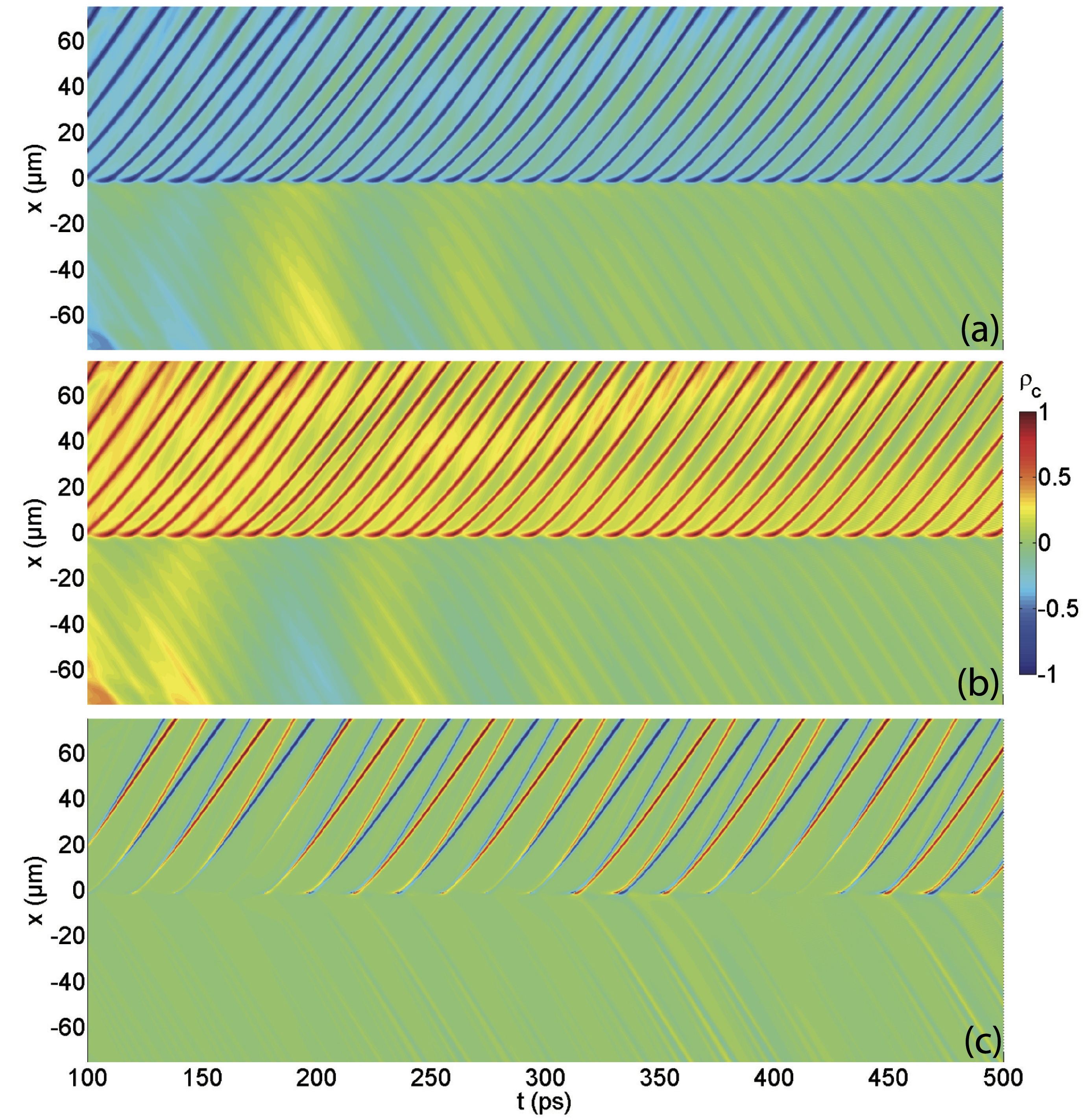}\\
\caption{(Color online) Control over the polarization of the trains. The colormap shows the degree of circular polarization $\rho_c$. (a) $\gamma_2/\gamma_1=0.90$, (b) $\gamma_2/\gamma_1=1.11$ and (c) $\gamma_2/\gamma_1=0.99$.}
\label{Fig4}
\end{figure}

\emph{Polarization control}.--
So far we have discussed a phenomenology that could be reproduced using a scalar condensate without any need for its spinor character. Indeed we have considered the ideal case of a perfectly linearly polarized condensate in its ground state with no polarization symmetry breaking namely $n_+(x)=n_-(x)$ for any $x$ position. The consequence is that the dark solitons formed in one spin component are perfectly overlapping with the ones in the other component and are behaving as scalar ones. However in real experimental situations the fluctuations brought by the structural disorder or the background noise can affect the linear polarization of the condensate leading to local inhomogeneities slightly breaking the polarization symmetry or the equivalence between the two spin components. As discussed in \cite{Monopole2} these fluctuations can lead to the separation of dark solitons in each component to form pairs of half-solitons \cite{HS,HSS}. As soon as they are split they will repel each other under the condition $\alpha_2<0$ and start to feel an effective magnetic force imposed by $H_x$ and be accelerated or slowed down depending on their linear polarization texture \cite{Monopole2}. This effect produced by local inhomogeneities or random processes would obviously be harmful to the formation of a deterministic spin signal. However as it was observed experimentally in Ref.\cite{Kammann}, using a polarized excitation laser can lead to the formation of a circularly/elliptically polarized condensate due to the long characteristic spin relaxation times of excitons. In Fig.\ref{Fig4}, we show results capitalizing on this effect to produce a useful spin signal.

We have modeled a slightly elliptically polarized nonresonant pump introducing two different reservoir/condensate transfer rates $\gamma_1$ and $\gamma_2$ in Eqs.(\ref{psi1},\ref{psi2}). We have adjusted the ratio $\gamma_2/\gamma_1$ to 0.90, 1.11 and 0.99 in panels (a), (b) and (c) respectively. The colormap shows the degree of circular polarization $\rho_c(x,t)=(n_+-n_-)/(n_++n_-)$ of the polariton emission. We see that the weak ensuing density imbalance between the two spin components of the condensate leads to a well defined polarization symmetry breaking inducing either the formation of trains of pairs of half-solitons [panel (c)] or trains of half-solitons with a well defined polarization for larger imbalances [panels (a) and (b)]. It means that not only the frequency of the trains can be finely tuned but also their polarization by variation of the input polarization. It provides another degree of freedom to code information.

\emph{Conclusions}.--
We have shown the strong potential of microcavities for high speed optical signal modulation for information coding. Our proposal involves the all-optical or electric control of dark soliton trains within a realistic and experimentally trivial scheme. We have demonstrated the possibility to tune both the train frequency and its polarization. The present concept could play a central role at the heart of future high speed polariton circuits within the rapidly expanding field of spin-optronics.

\emph{Acknowledgements}.--
F.P. acknowledges financial support by the UK Engineering and Physical Sciences Research Council (EPSRC) grant EP/H023348/1 for the University of Cambridge Centre for Doctoral Training, the Cambridge Centre for Analysis and a KAUST grant. H. F. acknowledges financial support from NCCR Quantum Photonics (NCCR QP), research instrument of the Swiss National Science Foundation (SNSF). We thank F. Manni and N. Berloff for fruitful discussions.

 \end{document}